\documentclass[twocolumn,showpacs,preprintnumbers]{revtex4}

\usepackage{graphicx}
\usepackage{dcolumn}
\usepackage{amsmath}
\usepackage{amssymb}

\begin{document}

\preprint{}
\title{Tunneling between Two Helical Superconductors via Majorana Edge Channels
}
\author{Yasuhiro Asano$^{1}$, Yukio Tanaka$^{2}$, and Naoto Nagaosa$^{3,4}$}
\affiliation{
$^1$Department of Applied Physics \& Center for Topological Science and Technology, 
Hokkaido University, Sapporo, 060-8628, Japan. \\ 
$^2$Department of Applied Physics, Nagoya University, Nagoya, 464-8603, Japan. \\ 
$^3$ Department of Applied Physics, University of Tokyo, Tokyo, 113-8656, Japan. \\
$^4$ Cross Correlated Materials Research Group (CMRG), ASI, RIKEN, WAKO 351-0198, Japan. 
}
\date{\today}

\begin{abstract}
We discuss electric transport through a point contact 
which bridges Majorana fermion modes appearing at edges of two helical superconductors.
The contents focus on effects of interference and interaction unique to the Majorana fermions
and role of spin-orbit interaction (SOI). 
Besides the Josephson current, the quasi-particle conductance depends sensitively on 
phase difference and relative helicity between the two superconductors.  
The interaction among the Majorana fermions causes the power-law temperature 
dependences of conductance for various tunneling channels. 
Especially, in the presence of SOI, the conductance always 
increases as the temperature is lowered. 
\end{abstract}

\maketitle



%

%




Fractionalization of electrons attracts recent intensive interest.
A chiral fermion at an edge of Quantum Hall system is an example of 
fractionalized electrons. A fully gapped bulk state spatially separates right-going 
and left-going chiral fermion, which leads to the absence the backward scattering
~\cite{Wen}.
The robustness against disturbances such as disorder and interaction
 is a common feature of the fractionalized states.
Therefore the fractionalized states are expected to have 
dissipationless feature which is a key property  
on application to quantum information processes~\cite{Nayak}.
Majorana fermion (MF) is another example of the fractionalized 
electron and has been recently discussed in the context of 
condensed matter physics~\cite{gogolin}. 
Since its field operator in real space satisfies a relation
$\gamma=\gamma^\dagger$, MF is often called 
real fermion and is regarded as $\textit{a half}$ (fraction) of 
a usual complex fermion. 

Superconductors and superfluids are the most promising  
candidates which host MF because the particle number is not a good quantum number
in these systems as required by the MF field.
Actually the existence of MF has been  
discussed in a vortex core or at an edge of the chiral $p+ip$ 
superconductor~\cite{Ivanov2001}, $^3$He and Bose-Einstein condensates~\cite{Volovik,Tewari}, 
at an interface between a superconductor and a topological 
insulator~\cite{Fu2008}, and
in a quantum Hall edge (state) with $\nu=5/2$~\cite{Read2000}.
Chiral fermion modes appear only when the time-reversal symmetry
$\mathcal{T}$ is broken \cite{Wen}. 
Under preserving $\mathcal{T}$-symmetry, the partner 
with the opposite chirality always coexists. In this case, the edge channels
are referred to as $\textit{helical}$. 
Along the edge of the two-dimensional quantum spin Hall systems, 
the helical fermions appear \cite{Helical}. 
Analogously, at the edge of the helical superconductors, the helical Majorana
fermions appear as the Andreev bound states~\cite{Furusakia,Qi,ABS}. 
Noncentrosymmetric superconductors with dominant spin-triplet 
$p$-wave order parameter are a realistic playground of helical 
Majorana fermions~\cite{Furusakia,Qi}. In addition, MF 
excitations are expected in topological superconductors~\cite{hor_arxiv_09}. 
Thus understanding of novel phenomena peculiar to MF is
highly desired~\cite{tanaka_prl_09}. 
Although all of recent developments have assumed non-interacting MF's, 
effects of interaction among MF's have been an important open question. 

  In this paper, we study theoretically low energy electric transport 
through the MF modes appearing at edges of 
 two helical superconductors with taking into account the 
interaction among MF's.  The model of four interacting MF modes
 can be mapped into the spinless Tomonaga-Luttinger
model by introducing two fictitious chiral fermions. 
This enables us to analyze low energy physical phenomena of a MF
by using the bosonization technique. 
 We show that the conductance depends sensitively on 
the relative helicity of two helical superconductors, 
the phase difference of two superconductors, 
 and the spin-orbit coupling at the point contact. 
These are the features of the interference effect unique to Majorana Josephson junctions.  
Our main results are summarized in Eqs.~(\ref{josephson}),
(\ref{sigmae}), (\ref{josephson2}) and (\ref{sigmao}).

\begin{figure}[tbh]
\begin{center}
\includegraphics[width=8.0cm]{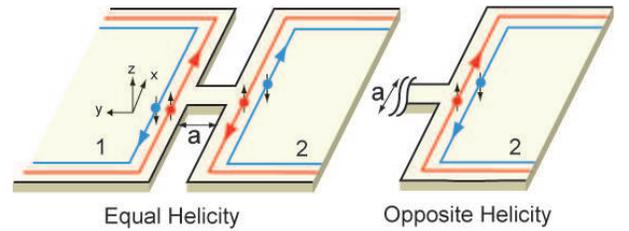}
\end{center}
\caption{
(Color online) 
Schematic picture of helical Majorana excitations at edges of two helical superconductors. 
}
\label{fig1}
\end{figure}

We consider the interacting helical edge channels~\cite{tao} as shown in Fig.1.
By using solutions of the Bogoliubov-de Gennes equation, Majorana fermions 
at the edge of the helical superconductors are described by~\cite{ABS}
\begin{align}
H_0=& -i v \sum_{j=1,2}\int d x \left[ \gamma_{Rj}(x) \partial_x \gamma_{Rj}(x)
\right.\nonumber\\  
& \left.   - \gamma_{Lj}(x) \partial_x \gamma_{Lj}(x)\right]
\end{align}
where $\gamma_\mu(x)$ for $\mu=(R1, L1, R2, L2)$ are the four species of 
Majorana fermion field satisfying the anti-commutation relation 
$\{ \gamma_{\mu}(x), \gamma_{\mu'}(x') \} = (1/2) \delta_{\mu,\mu'} \delta(x-x')$.
The electron operator is expressed in the low energy sector as
\begin{align}
\Psi_{j,\sigma}(x) =& e^{i\varphi_j/2} e^{i\varphi_\sigma} \chi \gamma_{j,\sigma}(x),  
\label{psidef}\\
\chi =& \left\{\begin{array}{cc} e^{i\pi/4} & (j,\sigma)=(1,\uparrow)\, \text{and}\,
 (2,\downarrow)\\
e^{-i\pi/4}& (j,\sigma)=(1,\downarrow)\, \text{and}\, (2,\uparrow),\end{array} \right.
\label{Psi}
\end{align}
where $\varphi_{j}$ is the phase of superconducting order parameter for 
the two superconductors indicated by $j=1$ and 2, 
$\sigma=\uparrow, \downarrow$ represents pseudospin index, and
$\varphi_\sigma$ is the relative phase of the pair potential for the two pseudospins.
In Eq.~(\ref{psidef}), we have considered 
 pair potential in two-dimensional $^3$He-B phase as an example of the helical edge state.
  As shown in Fig.~\ref{fig1}, the pseudospin index $\sigma=\uparrow, \downarrow$ is
related to the moving direction of the chiral Majorana fermions ($R, L$), which  
are basically determined by the pair potential and the spin-orbit interaction (SOI) 
in bulk region. 

In addition to $H_0$, we consider the two terms 
in the Hamiltonian. At first, the interaction $H_{\textrm{int.}}$ comes from the 
electron-electron interaction as given by 
$H_{\textrm{el-int.}}=\int dx \int d x'  C^\dagger_\alpha (x) C^\dagger_{\beta}(x')
V(x-x') C_\beta(x') C_{\alpha}(x)/2$, where $C_{\alpha(\beta)}$ is the electron operator
with $\alpha$ and $\beta$ labeling the electron spin. 
The original electron spin $\alpha$ is expressed by the linear 
combination of pseudospin $\sigma$ in the presence of the SOI.
From the fact that $\Psi_{i \sigma}^\dagger \Psi_{i \sigma} \propto 
(\gamma_{i \sigma})^2 = \textrm{const.}$, and
assuming that the overlap of 
the wavefunctions between the two edges is negligible, the only remaining 
interactions derived from $H_{\textrm{el-int.}}$ in low energy are
\begin{eqnarray}
H_1 &=& U_1 \int dx [ \Psi^\dagger_{1,\uparrow} \Psi_{1,\downarrow} 
\Psi^\dagger_{2,\uparrow} \Psi_{2,\downarrow} + h.c.] 
\nonumber \\
H_2 &=& U_2 \int dx [ \Psi^\dagger_{1,\uparrow} \Psi_{1,\downarrow} 
\Psi_{2,\uparrow} \Psi^\dagger_{2,\downarrow} + h.c.].  
\label{interactions}
\end{eqnarray}
Note here that more than four species of Majorana fermions and SOI are indispensable 
to having effective interaction~\cite{Ho} 
and that interaction terms including spatial derivatives are irrelevant in the 
low energy limit. 
Expressing Eqs.~(\ref{interactions}) by Eq.~(\ref{Psi}), we 
obtain
\begin{equation}    
H_{\textrm{int.}}= g \int dx   \gamma_{R1}(x) \gamma_{R2}(x) \gamma_{L2}(x) \gamma_{L1}(x),
\end{equation}
with $g= - 2 U_1 \cos(2 \varphi_s) + 2 U_2$.

The last term is the tunneling Hamiltonian between the two edges represented by
\begin{align}
H_T=& -ta \sum_{\sigma,\sigma'} \left[ 
\Psi_{1,\sigma}^\dagger(0)  
\left\{ \sigma_0 + i \boldsymbol{\lambda}\cdot\boldsymbol{\sigma} 
\right\}_{\sigma,\sigma'} \Psi_{2,\sigma'}(0) \right. \nonumber\\
&+  \left.\Psi_{2,\sigma}^\dagger(0) 
\left\{\sigma_0 -i \boldsymbol{\lambda}\cdot\boldsymbol{\sigma} \right\}_{\sigma,\sigma'} 
\Psi_{1,\sigma'}(0) \right],\label{ht}\\
=&2ita \left[ \cos(\varphi/2) A_- + \lambda_3 \sin(\varphi/2) A_+ \right. \nonumber\\
&-\cos(\varphi/2)\lambda_- B_+ \nonumber+\left.\sin(\varphi/2)\lambda_+ B_-
\right], \label{ht1}\\
A_{\pm}=&\gamma_{1,\uparrow}\gamma_{2,\uparrow} \pm \gamma_{1,\downarrow}\gamma_{2,\downarrow}, \,
B_{\pm}=\gamma_{1,\uparrow}\gamma_{2,\downarrow} \pm \gamma_{1,\downarrow}\gamma_{2,\uparrow},
\end{align}
with $\lambda_-= \lambda_1 \cos\varphi_{s}-\lambda_2\sin\varphi_{s}$ and 
$\lambda_+ = \lambda_1 \sin\varphi_{s}+\lambda_2\cos\varphi_{s}$,
 where $\varphi=\varphi_1-\varphi_2$ is the macroscopic phase difference, 
$\varphi_s=\varphi_\uparrow-\varphi_\downarrow$ is the difference 
in the spin-dependent phases,
$\boldsymbol{\sigma}=(\sigma_1, \sigma_2, \sigma_3)$ are the 
Pauli matrices, and $\sigma_0$ is the $2\times 2$ unit matrix.
The width and the length of a point contact is indicated by $a$.
In Eq.~(\ref{ht}), we consider the SOI at the point 
contact described by 
$\boldsymbol{\lambda}=(\lambda_1, \lambda_2, \lambda_3)$. 
By applying electric fields at the point contact, it is possible to induce
the Rashba-type SOI as 
$\boldsymbol{\lambda} = g_{so}(-E_z, 0, E_x)$, where 
$g_{so}$ is a coupling constant,
$E_x$ and $E_z$ correspond to the potential gradient in the $x$ and $z$ direction, respectively. 
When the Dresselhous-type SOI can be introduced, 
$\lambda_2$ also becomes non-zero value.
The operator of the electric current is calculated from the equation
$J=e \partial_t \sum_\sigma \int dx \Psi^\dagger_{1,\sigma}(x)\Psi_{1,\sigma}(x)$. 
We find that a relation $J(\varphi)=eH_T(\varphi+\pi)$ always holds.
We assume that the pseudospin of the right(left)-mover is $\uparrow$ ($\downarrow$) 
at the edge 1 as shown in Fig.~\ref{fig1}. 
Therefore we choose $\gamma_{1,\uparrow}=\gamma_{R1}$ and 
$\gamma_{1,\downarrow}=\gamma_{L1}$. 
At the edge 2, we choose $\gamma_{2,\uparrow}=\gamma_{L2}$ and 
$\gamma_{2,\downarrow}=\gamma_{R2}$ for the equal helicity configuration, 
and $\gamma_{2,\uparrow}=\gamma_{R2}$ and $\gamma_{2,\downarrow}=\gamma_{L2}$ 
for the opposite helicity configuration. 

To analyze the Hamiltonian, we introduce a complex fermion field by
$\psi_A(x) = \gamma_{A1}(x) + i \gamma_{A2}(x)$ and $\psi_A^\dagger(x) =
 \gamma_{A1}(x) - i \gamma_{A2}(x)$ with $A=R$ and $L$. 
These operators satisfy the usual fermion anti-commutation relations:
$\{ \psi_{A}(x), \psi_{A'}(x') \} =0$ and 
$\{ \psi_{A}(x), \psi_{A'}^\dagger(x') \} = \delta_{A,A'} \delta(x-x')$.
We rewrite the Hamiltonian $H_0 + H_{\textrm{int.}}$ in terms of these complex fermion operators as
\begin{align}
H_0 +& H_{\textrm{int.}} = -iv \int d x \left[ \psi_{R}^\dagger(x) \partial_x \psi_{R}(x) 
-  \psi_{L}^\dagger(x) \partial_x \psi_{L}(x) \right. \nonumber\\
  &+ \left. \frac{g}{4} \psi_{R}^\dagger(x) \psi_{R}(x)
 \psi_{L}^\dagger(x) \psi_{L}(x) \right] + \textrm{const.}.
\label{eq:H}
\end{align}
This Hamiltonian is exactly that of the massless Tomonaga-Luttinger model. 
It is extremely simple as compared to that of 
interacting helical edge fermion systems~\cite{Wu2006}. 
The combining two Majorana fermions corresponds to 
a spinless chiral fermion. Thus a chiral
Majorana fermion is considered as a quarter fraction
of a usual spinless fermion.
By applying the standard bosonization technique~\cite{Giamarchi,Solyom}, 
the complex fermion fields are represented by boson fields as,
\begin{align}
&\frac{\partial_x}{2\pi}\phi_{L(R)}(x)=:\psi_{L(R)}^\dagger(x) \psi_{L(R)}^{ }(x):,\\ 
&\phi(x)=\phi_R(x)+\phi_L(x),\, \theta(x)=\phi_R(x)-\phi_L(x),\\
&\psi_R(x)= \frac{\eta_R}{\sqrt{2\pi \alpha_0}} 
\exp\left[ \frac{i}{2}\left\{ \phi(x) + \theta(x) \right\}\right],\\
&\psi_L(x)= \frac{\eta_L}{\sqrt{2\pi \alpha_0}} 
\exp\left[ \frac{i}{2}\left\{ - \phi(x) + \theta(x) \right\}\right], 
\end{align}
where $\eta_R$ and $\eta_L$ are the Klein factor.
Eq.~(\ref{eq:H}) is then transformed into 
\begin{align}
H_0 &+ H_{\textrm{int.}} =\frac{\tilde{v}}{8\pi} \int \!\! dx 
\frac{ \left\{\partial_x \phi(x)\right\}^2}{K} + 
K \left\{\partial_x \theta(x)\right\}^2
,\label{hk}\\
\tilde{v}=&v \sqrt{1-\left(\frac{g}{8\pi v}\right)^2}, 
K= \sqrt{ \frac{1-g/(8 \pi v)}{1+g/(8 \pi v)} }.
\end{align}

We first discuss the tunneling effect in the \textit{equal helicity configuration}.
By using the bosonization technique, the tunneling Hamiltonian becomes
\begin{align}
&H_{T}= \frac{ta}{\pi} \left[ \frac{\lambda_+}{2}\sin\left[\frac{\varphi}{2}\right] \partial_x\theta(x)
- \frac{\lambda_-}{2}\cos\left[\frac{\varphi}{2}\right] \partial_x\phi(x)
  \right]_{x=0} \nonumber\\
&+\frac{i\eta_R\eta_L ta }{\pi \alpha_0}\!\!
\left[ \cos\!\left[\frac{\varphi}{2}\right] \sin\theta(0) - \lambda_3\sin\!\left[\frac{\varphi}{2}\right] \sin\phi(0) \right]. \label{hte}
\end{align}
In Eq.~(\ref{hte}), the terms including $\partial_x\theta(x)$ and $\partial_x\phi(x)$ 
represent the forward tunneling process: hopping of left(right)-mover to left(right)-mover.
On the other hand, the terms including $\sin\theta(0)$ and $\sin\phi(0)$ represent
the backward tunneling: hopping of left(right)-mover to right(left)-mover.
Before turning into the conductance, the Josephson current should be clarified.
Within the second order perturbation expansion, we find 
\begin{align}
 J =&  e \Delta \left[\frac{at}{\pi v}\right]^2 \!\!\!\sin\varphi \left[
1-\lambda_3^2 - \frac{1}{K} \lambda_+^2 +  K \lambda_-^2 \right], \label{josephson}
\end{align}
where we have used $(\eta_R\eta_L)^2=-1$, $(\alpha_0)^{-1}=k_F \equiv \Delta/v$, and 
$\Delta$ is the amplitude of pair potential at sufficiently 
low temperature $T\ll T_c$ with $T_c$ being superconducting transition temperature.
The ground state of junction is at $\varphi=0$ for $\boldsymbol{\lambda}=0$~\cite{kwon}. 
 Eq.~(\ref{josephson}) does not recover a usual relation
 $J\propto (1+\boldsymbol{\lambda}^2)$ expected in $s$-wave Josephson junction
even in the absence of interaction, (i.e., $K=1$). 
This is a characteristic feature of Majorana fermion excitation. 

On the basis of the linear response theory, we calculate DC conductance of the point contact
using the standard Kubo formula,
$\sigma= - \lim_{\omega \to 0^+} [Q^R(\omega)-Q^R(0)]/(i\omega)$, 
where the correlation function is obtained by
$Q^R(\omega)=Q(i\omega_n \to \omega+i\delta)$ with
$Q(\omega_n)= - \int_0^{1/T}d\tau e^{i\omega_n\tau} \left\langle J(\tau) J(0) \right\rangle$, 
$\tau$ is the imaginary time, and $\omega_n$ is the Matsubara frequency.
The following four terms contribute to $Q(\omega_n)$,
\begin{align}
&\langle J(\tau) J(0) \rangle = \left\langle 
F_\theta^2 \partial_x \theta(\tau) \partial_x \theta(0) +
F_\phi^2 \partial_x \phi(\tau) \partial_x \phi(0)\right\rangle_{x=0}\nonumber\\
&+\left\langle B_\theta^2 \sin\theta(\tau)\sin\theta(0) 
+ B_\phi^2 \sin\phi(\tau)\sin\phi(0) \right\rangle_{x=0},
\end{align} 
where $F_\theta=\cos(\varphi/2)\lambda_+/2$, $F_\phi=\sin(\varphi/2)\lambda_-/2$, 
$B_\theta=\sin(\varphi/2)/\alpha_0$ and $B_\phi=\cos(\varphi/2)\lambda_3/\alpha_0$.
After deriving the effective 
action for $\theta(0,\tau)$ and $\phi(0,\tau)$, we obtain the 
scaling equations by following Refs.~\cite{FisherZwerger,Furusaki,KaneFisher},
\begin{align}
\frac{d B_\theta^2(l)}{dl}=& 2\left(1-\frac{1}{K}\right)  B_\theta^2(l),\\
\frac{d B_\phi^2(l)}{dl}=& 2\left(1-K\right)  B_\phi^2(l),
\end{align}
where $\mu = \Lambda e^{-l}$ gives a boundary between high- and low-frequency 
and $\Lambda$ is the high-frequency cut-off. 
It is concluded that $B_\theta$ and $B_\phi$ are relevant 
for $K>1$ ($g<0$) and  $K<1$ ($g>0$), respectively.
\begin{table}
\begin{ruledtabular}
\begin{tabular}{cccc}
\null & \null& $\boldsymbol{\lambda}=0$  & $\boldsymbol{\lambda}\neq 0$\\
\colrule
{Equal helicity} & & & \\
$\varphi=0$ & $K=1$ & 0  & const. \\
\null & $K<1$ & 0  & $T^{2K-2}$ \\
\null & $K>1$ & 0  &  const. \\
$\varphi\neq 0$& $K=1$ & const. &  const.\\
\null & $K<1$ & $T^{2/K-2}\to 0$ &  $T^{2K-2}$\\
\null & $K>1$ & $T^{2/K-2}$ &  $T^{2/K-2}$\\
\colrule
{Opposite helicity} & & & \\
$\varphi=0$ & $K=1$ & 0 & const. \\
\null& $K<1$ & 0 & const. \\
\null & $K>1$ & 0 & $T^{2/K-2}$\\
$\varphi\neq 0$& $K=1$ & const. &  const.\\
\null & $K<1$ & const. & $T^{2K-2}$\\
\null & $K>1$ & const. & $T^{2/K-2}$\\
\end{tabular}
\end{ruledtabular}
\caption{The temperature dependence of the most dominant terms in conductances
given in Eqs.~(\ref{sigmae}) and (\ref{sigmao}) at low temperature, where 
 $\varphi$ is the phase difference between the two helical superconductors,
$\boldsymbol{\lambda}$ represents spin orbit interaction at a point contact, and
 'const.' means the conductance independent of temperature.
Here $K=1 (g=0)$ represents no interacting case, while 
$K<1 (g>0)$ and $K>1 (g<0)$ are the interacting cases. 
The conductance depends on the relative helicity of the two superconductors:
the equal helicity configuration (upper column) and the opposite one (lower column).
}
\label{table1}
\end{table}

The forward tunneling terms are calculated to be
\begin{align}
&\langle \partial_x\theta(x,\tau)\partial_x\theta(x,0)|_{x=x_0} \rangle 
= \frac{1}{K}X(\tau)\\
&\langle \partial_x\phi(x,\tau)\partial_x\phi(x,0)|_{x=x_0} \rangle 
= KX(\tau),\\
&X(\tau)=\frac{8k_F}{v} \delta(\tau) - \frac{4\pi}{v^2} 
T \sum_{\omega_n}e^{-i\omega_n\tau}|\omega_n|.
\end{align}
Finally we reach the DC conductance for the \textit{equal helicity configuration}
\begin{align}
&\frac{\sigma}{G_0} =  \pi\frac{\lambda_+^2}{K}\cos^2\left(\frac{\varphi}{2}\right) 
+ { \sin^2\left(\frac{\varphi}{2}\right)} D_\theta\left(\frac{T}{T_0}\right)^{2/K-2}\nonumber\\
&+ \pi \lambda_-^2 K \sin^2\left(\frac{\varphi}{2}\right)
 +  
{ \lambda^2_3 \cos^2\left(\frac{\varphi}{2}\right)} D_{\phi}\left(\frac{T}{T_0}\right)^{2K-2},
\label{sigmae}
\end{align}
where $G_0=(tae/\pi v)^2$ and 
$D_\mathcal{A}(T_0)= \Delta_0^2 \int_0^{1/T_0} d\tau \tau \langle \sin\mathcal{A}(\tau) \sin\mathcal{A}(0) \rangle$ for $\mathcal{A}=\theta$ and $\phi$ are the correlation function at $T=T_0 < T_c$.
Remarkably, the conductance depends on the phase difference between the two superconductors.
This stems from a peculiar feature of Majorana fermion. The two terms in the tunneling 
Hamiltonian, which are Hermite conjugate to each other, turn into the same form 
due to the "real" nature of Majorana fermion giving the interference effect. 
At $\varphi=0$, the conductance vanishes 
in the absence of the SOI, (i.e., $\boldsymbol{\lambda}=\boldsymbol{0}$). 
In the presence of the SOI at $\varphi=0$, the last term is relevant 
for $g>0$ in addition to the first term. 
The first term is independent of $T$, whereas 
the last term increases with decreasing $T$ as $T^{2K-2}$ for $g>0$. 
For $\varphi \neq 0$, the second term is relevant for $g<0$ even in the absence of
the SOI. Finally for  $\varphi \neq 0$ and $\boldsymbol{\lambda}\neq 
\boldsymbol{0}$, all terms contribute to the conductance.
The argument above is summarized in Table~\ref{table1}. 

The total current through a Josephson junction is described by so called resistively and 
capacitively shunted junction model
\begin{align}
J = \frac{C}{2e} \frac{d^2 \varphi}{dt^2} + \frac{1}{2eR(\varphi)}\frac{d\varphi}{dt}
+ J_0 \sin(\varphi),
\end{align}
with $C$ being the capacitance of a junction. In the present junction, the resistance
$R=1/\sigma$ depends on $\varphi$. Thus Majorana fermion excitation would modify dynamics 
of a Josephson junction. The phase $\varphi$ may be determined self-consistently 
so that the current can be optimized. Such issue is a natural extension of this paper.

In the case of \textit{opposite helicity configuration}, we also obtain the Josephson current and 
the conductance as follows
\begin{align}
&\langle J\rangle =  e \Delta \left[\frac{at}{\pi v}\right]^2 \!\!\!\sin\varphi \left[
\frac{1}{K}-\lambda_3^2K - \lambda_+^2 +  \lambda_-^2 \right],\label{josephson2}
\end{align}
\begin{align}
&\frac{\sigma}{G_0} = \pi \frac{\sin^2(\varphi/2)}{K} + { \lambda_+^2 \cos^2\left(\frac{\varphi}{2}\right)} D_\theta\left(\frac{T}{T_0}\right)^{2/K-2} \nonumber\\
&+ \pi \lambda_3K \cos^2(\varphi/2)+  
{ \lambda^2_- \sin^2\left(\frac{\varphi}{2}\right)} D_{\phi}\left(\frac{T}{T_0}\right)^{2K-2}.
\label{sigmao}
\end{align}
For $\boldsymbol{\lambda}=0$, the conductance proportional to $\sin^2(\varphi/2)$ is 
independent of temperature, which is in sharp contrast to that in the equal helicity case given 
in Eq.~(\ref{sigmae}). The behaviors of the conductance $\sigma$ are summarized in 
the Table~\ref{table1}.

In summary, we have studied electric transport through a point contact 
which connects Majorana fermion modes appearing at the edges of two helical superconductors
by taking into account the interaction among Majorana fermions and the spin-orbit interaction
(SOI) at a point contact. 
By introducing a fictitious fermion consisting of two Majorana
fermions, the Majorana fermion model is transformed into 
the Tomonaga-Luttinger model. The application of the standard 
bosonization technique enables to analyze low energy physical 
phenomena of a Majorana fermion. It is found that several novel 
features appear due to the Majorana fermions such 
as (i) the conductance is sensitive to the phase difference 
of two superconductors, 
(ii) tunneling with SOI gives quite distinct 
behavior from that without SOI, 
(iii) the transport phenomena depend on relative helicity of two superconductors
as shown in Eqs.~(\ref{sigmae}) and (\ref{sigmao}), 
and (iv) the interactions leads to the power-law temperature/voltage dependences.



\end{document}